# Using Bayesian Modelling to Predict Software Incidents

**Chris Hobbs, Waqar Ahmed**

BlackBerry QNX

Ottawa, Canada

**Abstract** *Traditionally, fault- or event-tree analyses or FMEAs have been used to estimate the probability of a safety-critical device creating a dangerous condition. However, these analysis techniques are less effective for systems primarily reliant on software, and are perhaps least effective in Safety of the Intended Functionality (SOTIF) environments, where the failure or dangerous situation occurs even though all components behaved as designed. This paper describes an approach we are considering at BlackBerry QNX: using Bayesian Belief Networks to predict defects in embedded software, and reports on early results from our research.*

## 1  Introduction

Particularly when used in critical applications, software for embedded systems is generally developed so as to mitigate faults which can escalate to system failure. Despite verification analysis designed to identify and remove faults, software bugs still cause failures in the field. It is important to be able to predict the number of failures that are likely to occur in the field.

## 1.1  The Problem

Even in a very simple software system running on a simple hardware platform, software failures tend to occur at random. Software does not wear out as mechanical systems do, but it can fail randomly, as (IEC 61513 2011) explains: *Since the triggers which activate software faults are encountered at random during system operation, software failures also occur randomly*. With modern (semi-)autonomous systems these triggers may have a long-tail distribution, see (Koopman 2018), and be effectively impossible to predict when the product is being designed. (Zhao et al.

2019) illustrates the basic impossibility of providing sufficient testing for an autonomous road vehicle to ensure that the vehicles are safe for public use.

Some standards, such as (ISO 26262 2018), do not accept the random nature of software failure and recommend only qualitative software failure analysis. Nonetheless, the quantitative analysis of software failure in complex modern systems, such as those in autonomous vehicles, is an essential approach to predicting system dependability early in the design stage and when we must decide if a product is sufficiently safe for shipment.

Software faults (bugs) which cause these occasional, randomly-distributed failures are known as Heisenbugs. The opinion that a modern software product can be free of Heisenbugs is unrealistic and, in the case of a safety-related product, dangerous. In short, before a product is shipped for use in a safety-critical application, it is essential that safety and verification specialists estimate the effect Heisenbugs may have on the product's operation.

## 1.2   The Traditional Approach

BlackBerry QNX has employed various dependability analysis techniques, including Failure Mode and Effect Analysis (FMEA), and Fault Tree (FT) (IEC 61025 (2006)) to estimate the effect of Heisenbugs on its software. Both these approaches require failure rates to predict the dependability of the software system. For commonly-deployed software such as an operating system (OS), over time the system's field usage (hundreds of millions of hours) grows rapidly compared to the number of bugs reported from the field, so it becomes unrealistic to use failure rates to predict the dependability of new releases.

## 1.3   The Approach Being Explored

(Fenton et al. 2008) present a novel approach to predicting the number of software defects; this approach is based on the quality and management of the software development processes and does not require analysis of failure rates to estimate software dependability. Using a Bayesian Network (BN) technique, they propose a defect prediction model with nodes representing various software quality metrics, such as development quality, verification quality and problem complexity. The prime benefit of BN-based defect model is that it allows incorporation of soft evidence later in the model development, which permits us to refine our defect predictions.

However, (Fenton et al, 2008) states: *"As it stands, the model will be less relevant for safety critical software or core algorithmic software where very few post-release defects can be tolerated, and so where few defects will be found later in testing phases or in operational usage."* One purpose of the investigation reported in this paper was to test whether, by extending the model to include additional metrics, the model could be used for safety-critical software.

## 1.4   This Paper

This paper begins by describing the traditional approach BlackBerry QNX uses to estimate defect rates and enumerates some of its weaknesses (Section 2). It then expands on the BN approach with which BlackBerry QNX has been experimenting (Section 3), and summarises the results of these experiments (Section 4).

## 2   Background

Since 2010, BlackBerry QNX has used Bayesian Fault Tree (BFT) analysis for conducting defect prediction of its products, and in particular its operating system products.

## 2.1   A Conventional Fault Tree

An OS kernel is an event handler (i.e. it reacts to events); these events can be interrupts, exceptions or kernel calls. If no events occur, the kernel does not execute. This fact makes the top-level failure modes easy to identify because there are only a few ways in which an event handler can fail to fulfil its tasks: (i) it can lose the event (e.g. not pass an interrupt to a thread because of lack of memory), (ii) it can handle the event incorrectly (e.g. schedule the wrong thread when an interrupt occurs), and (iii) it can corrupt its internal state while apparently handling the event correctly.

Figure 1 presents a snapshot of the top level of a classic FT. The coloured nodes indicate off-page connections to sub-trees. The complete FT occupies seven pages of a document and is not, therefore, included in this paper.

As well as providing more expressive conjunctions than a Boolean Fault Tree (e.g. Noisy-Or), not only can a BFT determine the system failure probability from component probabilities, but it can also analyse the posterior probability of any component node in the event that the system has failed. For instance, in Figure 1, we can assign a soft evidence value to the top-level node (System Failure) and determine the failure probabilities of particular causes from the lower-level nodes.

In order to perform the failure analysis, we extract the failure rates of individual components or groups of components from historical information; that is, the hours of use and the number of reported failures.

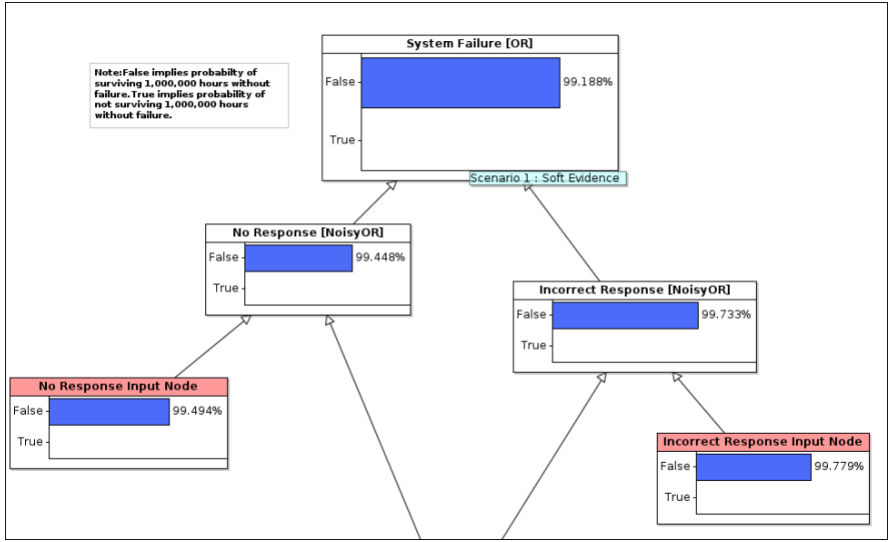

**Fig. 1.** Part of the top-level of a classic fault tree

We also make use of other failure computation techniques such as FMECA and Event-Tree analysis but found them inappropriate for software-based systems. In particular, when creating a FMECA, we often found it difficult to express the highly interrelated nature of software components and their dependencies on each other.

## 2.2   Weaknesses of the BFT Approach

Determining failure rates for the BFT analysis of software products mainly requires field usage history and the number of defects reported from the field. This approach has the following weaknesses:

1. The need for historical failure rates means that it can only be applied to those products with a field history, not to newly-developed products.
2. Not all products have easily classifiable failure modes.
3. Not all failures encountered in the field are reported. For example, a driver whose car infotainment system fails would probably just turn it off and on again; unless the problem were endemic, this failure would probably not be reported to the manufacturer. This means that we need to make an estimate of the proportion of problems being reported (often taken as 10%, but without justification for that value).
4. Changes that have been made to the product over time as a result of new features and bug fixes are not taken into account. (Ladkin 2013) points out that even changing one line of code results in a completely new piece of software and this is reflected in IEC 61508-3-1, and in (UL 4600 2019) which states: *Pitfall: Seemingly "small" changes to software and data are prone to causing potentially catastrophic item failures. The "size" and impact of a change cannot be assumed to be proportional to the number of lines of code or number of bytes of data changed.* Recognising this, we perform analyses immediately after a new release of a product to see whether there is a burst of reported bugs.

5. For a product such as the BlackBerry QNX OS, the usage hours are valid for those components that are in constant use in a customer's device; for example, the strict-priority scheduler, which is probably invoked thousands of times per second by every customer. The product also includes features, such as sporadic scheduling, that have limited use in the field. It is impossible to determine exactly which functions are actually being exercised in the field, how often they are exercised and for how long, so we should not apply the full "time-in-use" to all components in a product.
6. The approach is *a posteriori*; it does not provide useful information about the future, and cannot, therefore, support any decisions about when a new product is ready for release.
7. The approach provides metrics neither for the updating of processes and procedures, nor for assisting project management in estimating the resources and time required to complete the current or future projects.

## 3 Defect Prediction Modelling and Analysis: the Causal Approach

Defect prediction modelling is designed to answer the fundamental question: *If we ship the product today, how many defects are customers likely to encounter during the first T months of field usage?* For the deployment of a safety-critical product, the natural corollary to this question is: *Given the anticipated number of field failures, is it acceptable to ship the product now, or must we first undertake more verification and rework?*

(Shanmugam and Florence 2012) provides a high-level classification of many software reliability models, but many of the models they list are very old; perhaps the one most often used is that of (Littlewood and Verrall 1973). Additionally, some models are unsuitable for early prediction because they rely on analysis of the code, which means that they cannot be applied until at least some of the source code is available.

(Fenton et al. 2008) proposes a very different approach to answer the fundamental question. This approach is based on the assumption that the number of shipped defects is causally related to factors such as the complexity of the system being developed, the experience of the development team, and the quality of the verification performed. The remainder of this paper reports on our exploration of that approach at BlackBerry QNX.

(Fenton et al. 2008) describes the three-step approach normally used to build a model of software engineering:

1. Collect domain knowledge from history and expert knowledge.
2. Build a parametrised Bayesian model based on that knowledge without assigning values to the parameters.
3. Estimate values of the parameters through regression using historical information.

The experience described in (Fenton et al. 2008) is that, when building a causal model for 31 analysed developments, Step 3 was not required. Even without this step, the approach achieves remarkable accuracy in its predictions of the number of defects on the 31 projects.

Part of the exploration carried out by BlackBerry QNX was to determine whether Step 3 was required in its own environment. Our adaptation of the steps was as follows:

1. Collect domain knowledge of the factors likely to affect the field failure rate. We collected this knowledge from Project and Quality Managers of completed projects. Many of the factors they cited are probably common to all development organisations: the complexity of the product, the experience of the designers and implementers, etc. Other factors were specific to BlackBerry QNX: the maturity of the development procedures, the quality of the available tools, etc.
2. Build a Bayesian skeleton of the model incorporating the factors identified in Step 1. This model contains parameters associated with the combinations of factors with as yet unknown values; for example, whether the complexity of the product or the experience of the designers has the greater effect on the field failure rate?.
3. If necessary, use a combination of historical data and expert judgement to determine the value of the parameters in the model of Step 2.

BlackBerry QNX has copious data in various repositories covering its software development over almost two decades. These repositories include databases of all problems, whether found internally or reported from the field, all code changes with the identities of the person or persons who approved the code changes through the review process, detailed time-sheet records of the number of hours spent by different engineers on the project, detailed results of code, design and document reviews, estimates made at the "Gate" process which governs BlackBerry QNX developments, etc.

These data can be used to check for suspected correlations between, for example, code changes and failures, or code changes and further changes. For instance, are code changes made late on Friday afternoon, or changes approved by Bert and Ethel, likely to lead to further changes within a week? We have used these data to fill in the data required to conduct the BN analysis for some BlackBerry QNX software projects.

## 3.1  Building the model

The model proposed in (Fenton et al. 2008) is relatively simple, but it does contain one subtlety: if the field usage of a product is low, then the failure rate in the field will also be low. In the extreme case, if the product is not used at all, then no failures will occur! This is a useful observation and one which BlackBerry QNX incorporated into its model. Otherwise, the BlackBerry QNX model is very straightforward: the development group (inadvertently!) inserts bugs into the code, the verification group removes some of the bugs, and some of those that remain (residual bugs) inevitably cause failures in the field.

The Project and Quality Managers interviewed agreed that, apart from the amount of field use, the primary determiners of failure rates were verification quality, development quality and problem complexity, each of which we constructed as a significant sub-tree in our Bayesian tree (see Figure 2).

We then broke down each sub-tree, in some cases into further sub-trees, which we then also expanded. Anticipating that we would have to elicit information about shipped products from participants in those projects, we addressed advice given in (O'Hagan et al. 2006), including repeating some of the experiments described in that book. We realised that the definition of terms was very important and that we could not rely on different engineers consistently labelling the same characteristic as "high", "medium" or "low". This meant that we tried to make definitions as non-subjective as possible. See the discussion below on whether this approach is effective. For example, we accepted that part of the verification would be performed by dynamic testing and attempted to define the quality of that testing as follows:

- **Very High**: The testing is planned and executed as risk-based testing, as described in ISO 29119, and the risk to be mitigated is clearly defined before test case preparation starts. Testing is almost exclusively automated, making regression testing particularly effective.
- **High**: Requirements-based (rather than risk-based) testing is performed, with two-way tracing between requirements and test cases.
- **Medium**: Requirements-based testing is performed, without the requirements being traced explicitly to test cases.
- **Low**: There is no methodology other than error-guessing in the selection of test cases, and no coverage data collected during the tests.
- **Very Low**: The testing consists of ad hoc test cases executed manually by members of the verification team.

We stressed to those being interviewed that "soft" values were acceptable: for example 80% High, 20% Medium.

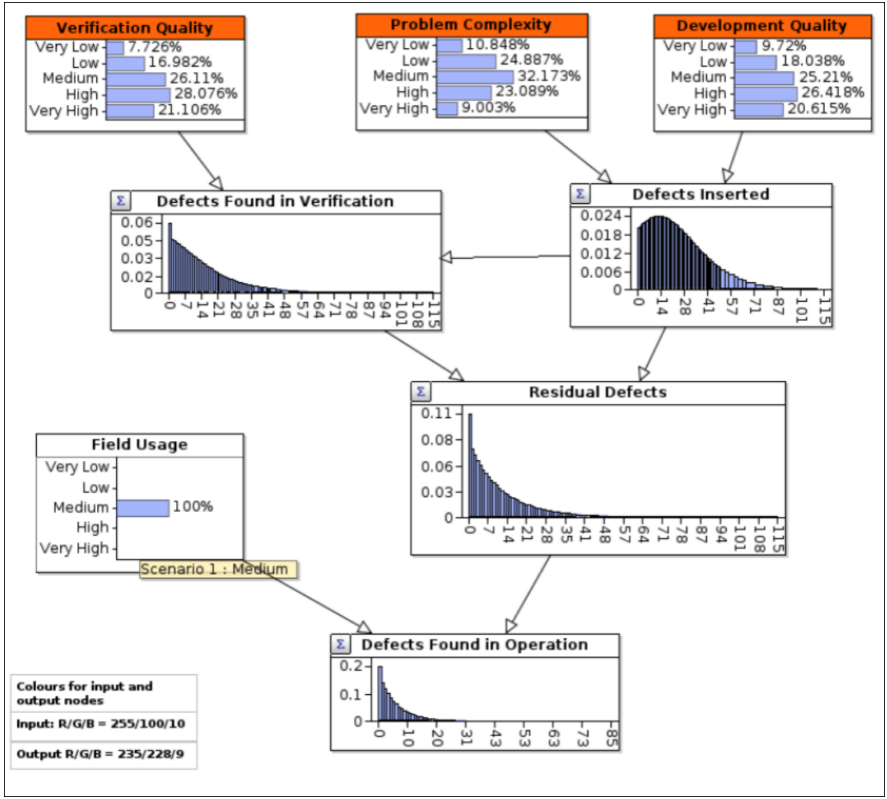

**Fig. 2.** Top-Level Model

## 3.2    Calibrating the model

We initially scaled the model against a small number of shipped BlackBerry QNX products, labelled in what follows as projects A, B and C; as our research progresses, we will expand this sample to other projects. Table 1 presents the characteristics of these projects. Each project was developed under the same Quality Management System, but with no common development engineers.

Project A was a redesign of an existing product, but with very few of the artefacts from the previous project brought forward.

Project B was the smallest of the three projects and the development team was very familiar with the application area, but, at the time of writing, the product had not been in the field for a complete year.

Project C was a complex development targeting a new area and not initially intended for safety certification, although various activities, such as Hazard and Risk Analyses, were performed to assist with future certification.

**Table 1.** Project Characteristics

| Project | Type of Development | Safety Certified? |
|---|---|---|
| A | Application (rather than OS) level development | Yes |
| B | Relatively small, very focused project to meet a published standard; the standard provided very clear and stable requirements. | Yes |
| C | Large development at the level of the processor hardware. | To be certified later |

## 4    Problems Encountered

While building the causal model for these projects, a number of problems arose that we had to address.

### 4.1    Objective or subjective questions?

We originally assumed that objective questions would give us the more useful values. For example, when assessing the quality of project management, rather than asking the subjective question, "How good was Project Management?", we defined specific criteria ranging from

- **Very High**: The project manager assigned to the project is certified as a Project Management Professional (PMP), and project resource estimation is performed using an industry-standard technique such as MODIST based on metrics from previous projects, with regular collection of statistics to monitor the project's progress, at least weekly.

through **High, Medium** and **Low** to

- **Very Low:** There is no independent project manager assigned to the project, and planning and monitoring are performed by the development manager.

In contrast, (Fenton et al. 2008) makes use of subjective questions, typically of the form *"How would you rate the quality of ...?"*

It is certainly true that a project manager having PMP qualifications does not guarantee a well-managed project, and asking the members of the project team *"How would you rate the quality of the project management?"* might lead to more accurate values. This question is subject to further study, which we are undertaking.

## 4.2   The impact of reused code

Reusing design and code from a previous project is a double-edged sword: it reduces the amount of new work, but it carries forward the as-yet undetected defects from the source project. It also assumes that all the constraints of the source project are clearly listed (as they were not, for example, in the code the caused the catastophic failure of the initial Ariane 5 launch). Project B in the study had made significant use of reused code, but this code originated as a reference implementation from an external standards body. As this code had been published and analysed by several companies, we decided that its net effect was to reduce, rather than increase, the complexity of the project.

## 4.3   The period over which to measure field-reported bugs

BlackBerry QNX provides components (operating systems, hypervisors, middleware) that its customers include in shipped product. This means that there is a delay between the BlackBerry QNX component being declared "generally available" and it actually appearing in the field: the product developer must incorporate the component into the product, verify it and ramp up production. The number of reported bugs during the intermediate period between the component being released and its appearing in production volume is not typical: the component is put under significant stress during the product developer's verification, but it is not released in large volume. Once it is in the field in perhaps a million cars, the product will be in use 730,000,000 hours per year, but under less stress than when being verified by the product developer.

This situation means that, unlike the projects listed in (Fenton et al. 2008), there are three distinct periods during which bugs are being reported: in house verification of the component, verification by the integrator, and field operation. Reasoning that during both of these periods our components are running in environments outside our control, we decided to combine the last two of these periods.

## 4.4  The number of reported bugs

In the 31 projects analysed in (Fenton et al. 2008), the mean number of defects per project found during the study period was 445, with one project reporting 1,906 defects. Although we did not take all the projects included in our study through ISO 26262 or IEC 61508 certification, the number of reported defects was much lower than this, and in one case it was zero. The output range being much tighter, this makes the accuracy of the model harder to assess in our case. (Fenton et al. 2008) acknowledges this difficulty in the quotation listed in section 1.3 above.

In general, the BlackBerry QNX software meets both of the criteria in the quotation: it is certified for use in safety critical systems and it provides core algorithms for thread scheduling, memory management, etc. However, even for software that meets these criteria, the increasing complexity of software has led to an increase in Heisenbugs which are not detected in verification when this is carried out through testing rather than formal analysis. We therefore added a node to the network to identify the type of verification carried out.

One other question which arose during the evaluation was whether all bugs are created equal. Should a bug that reports a misspelling in a displayed error string and a bug that reports a complete system failure each count as 1, or should there be a threshold below which a bug should not be counted? In this investigation we counted all bugs, irrespective of their importance.

## 4.5  Assessing the "size" of a project

One key attribute of a project's complexity is its "size". (Fenton et al. 2008) describes an original intention to use function points as an estimate of the project's size, but this was found to be impractical because it is not a metric commonly used in industry.

KLoC is known to be an inadequate metric, so for our model we incorporated a weighted balance of KLoC with the complexity of the new functionality to get an "effective KLoC". We then combined this with the number of hours booked to the development project to get an estimate of the project "size" as shown in figure Fig. 3.

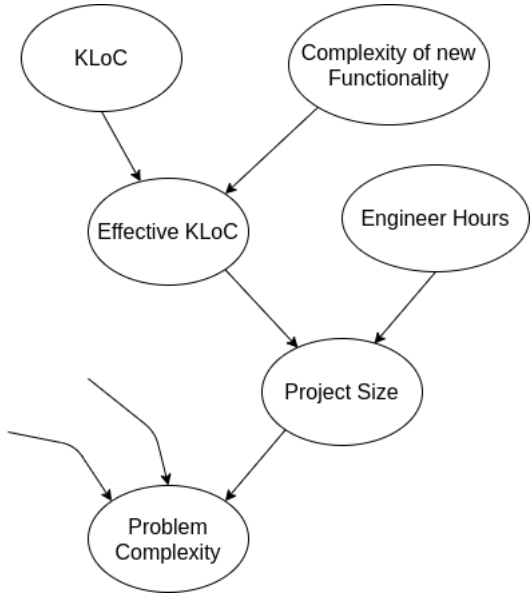

**Fig. 3.** Estimating project size

## 5   Results and Further Work

Table 2 provides the actual values (where known) and the numbers predicted by the BN for two of the key attributes during the forward pass through the network: the number of bugs reported during the verification period (either by the verification team itself, or by developers acting in the role of verifiers), and the number of bugs reported during the first year's deployment in the field. As can be seen, the actual and predicted values are disappointingly different. This means that step 3 in the process described in section 3 will need to be performed, calibrating the model against the actual data, and clearly more projects will need to be added to the analysis.

However, when the network is employed from effect to cause, by entering the actual number of faults found during verification, interesting dependencies emerge, particularly when sensitivity analyses are run. The model shows, for example, that whether the product is to be certified against ISO 26262 and IEC 61508 matters very little. This may be a reflexion on the quality of the underlying Quality Management System. Perhaps unsurprisingly, this analysis also shows a significant relationship between verification quality and number of defects likely to be found in the field.

However, (Fenton et al, 2008) does raise the question of whether finding a large number of defects during verification is a good or bad thing. If the verification quality is high, then finding many defects during the verification phase is presumably a good indication that most bugs are being removed before shipment. On the other hand, if the verification quality is low, then finding many bugs during verification is probably bad: indicating that the product is of poor quality.

**Table 2.** Numerical Results from the Model

|  | Defects found in verification | | Defects reported after shipment | |
| --- | --- | --- | --- | --- |
| Project | Actual Number | Number as predicted by the model | Actual Number | Number as predicted by the model |
| A | 76 | 109 | 8 | 11 |
| B | 10 | 74 | Not yet known | 8 |
| C | 195 | 136 | 24 | 97 |

Our project is very a much work-in-progress, but even the work currently completed has caused us to focus much more sharply on our project planning and management. (Yet et al, 2016) provides a model of the project planning and management process itself, including the ability to model initial and final budgets and timescales. At BlackBerry QNX the Project Management group has shown interest in the experiment described in this paper and (Yet et al, 2016) provides a potential extension to the current model to allow different project proposals to be compared to estimate ROI and project risk.

**Acknowledgments** The authors acknowledge the insights provided by Joe Bonner and Michael Lemke while the model was being built and thank Norman Fenton for providing useful information about the background to (Fenton et al, 2008).